\documentclass[english,aps,pra,twocolumn,superscriptaddress]{revtex4-1}
\usepackage[T1]{fontenc}
\usepackage{amsmath}
\usepackage{mathrsfs}
\usepackage{graphicx}
\usepackage{babel}
\usepackage{xcolor}

\begin{document}


\title{Spatial spin-wave modulator for quantum memory assisted adaptive measurements}


\author{Micha\l{} Lipka} 
\author{Adam Leszczy\'{n}ski}
\email[]{adam.leszczynski@fuw.edu.pl}
\author{Mateusz Mazelanik}
\author{Micha\l{} Parniak}
\author{Wojciech Wasilewski}

\affiliation{Centre for Quantum Optical Technologies, Centre of New Technologies, University of Warsaw, Banacha 2c, 02-097 Warsaw, Poland}
\affiliation{Faculty of Physics, University of Warsaw, Pasteura 5, 02-093 Warsaw, Poland}


\date{\today}

\begin{abstract}
Utilization of the spatial degree of freedom vastly enhances informational capacity of light at the cost of stringent requirements on the processing devices. Multi-mode quantum memories constitute a viable candidate for quantum and classical information processing; however, full utilization of the assets of high-dimensionality requires a flexible processing technique. We employ a spatially varying ac-Stark effect to perform arbitrary 1D phase modulation of a coherent spin-wave state stored in a wavevector-multiplexed quantum memory. 
A far-field and an interferometric near-field characterizations of the introduced phase profiles are presented. Additionally, coherence between temporally separated partial readouts of a single coherent spin-wave state is demonstrated, offering possible applications in adaptive measurements via conditional spin-wave modulation.
\end{abstract}


\maketitle

\section{Introduction}
An ability to shape and detect spatial and temporal structure of light lays at the foundations of modern quantum optical technologies \cite{Israel2017,Parniak2018_SR,Schwartz2013,Jachura2017,Jachura2018,Erkmen2010,Zhang2016}.
In particular utilization of the spatial degree of freedom offers a tremendous increase in the dimensionality of the accessible Hilbert space and information capacity. This asset has been vastly exploited in quantum optical communication \cite{Krenn2014,Babazadeh2017,Helwig2009,Usenko2014,Zhang2008}, quantum memories \cite{Parniak2017,Dabrowski2017_Optica},  and information processing \cite{Andersen2015}.
A parallel advent of highly multi-mode quantum memories \cite{Pu2017,Lan2009,Parniak2017,Vernaz-Gris2018}
equipped with the spatial processing capabilities \cite{Parniak2018,Mazelanik2018} further enhances the utilization of high dimensionality opening new avenues towards efficient realizations of communication and computation protocols e.g. enabling superadditive \cite{Klimek2016} or adaptive \cite{Wiseman2009} measurements. 
 
 
While several methods of information processing have been realized in quantum memories \cite{Monroe2013,Li2016}, thorough utilization of the spatial degree of freedom in highly-multi mode memories demands a flexible method analogous to spatial light modulators (SLM) widely exploited in the broad field of optics to modulate the amplitude or phase of light. In \cite{Parniak2018,Mazelanik2018} by employing spatially varying ac-Stark effect \cite{Leszczynski2018} simple few-mode operations have been demonstrated in the discrete domain on a coherent states of collective atomic excitations {\textendash} spin-waves as well as on single spin-wave Fock states, stored as a coherence between two metastable ground states in a cold $^{87}\mathrm{Rb}$ ensemble.

Here we characterize a spatial spin-wave modulator (SSM), allowing engineering of one dimensional (1D) spatial phase of spin-waves stored in a cold atomic memory. Phase modulation of a spin-wave state inherently transfers to the phase profile of light readout from the memory. SSM modulation offers flexibility of SLM modulators for light yet operates in the matter domain enabling long interaction times and opening broad possibilities for continuous domain quantum information processing in quantum memories. A full quantum information framework is thus built for spin waves in an analogy to electro-optic and dispersive manipulation of spectro-temporal degrees of freedom of photons \cite{Karpinski2016}.
  
In the far-field the SSM modulation enables shaping of the spatial intensity profile of the readout pulse and may be utilized for memory readout routing or mode matching e.g. for an enhanced coupling to an external photonic interface. With the ability to introduce parabolic phase profiles on a $\mu\mathrm{s}$ time scale, the SSM also provides a convenient method to dynamically switch between position {\textendash} momentum measurement bases expanding quantum information processing capabilities or aiding fundamental study of multi-dimensional entanglement
\cite{Dabrowski2017,Dabrowski2017_Optica,Dixon2012}.
Additionally, with the possibility of temporally sequencing or splitting the readout of a single coherent spin-wave state into several light pulses \cite{Hosseini2009,Reim2012,Cho2016}, an arbitrary phase modulation could serve in a feed-back loop for adaptive measurements e.g. enhancing the discrimination of non-orthogonal states of light \cite{Dolinar73,Cook2007,Assalini2011,Becerra2013,Hetet2015}.

Importantly, SSM modulation can be applied in a broad family of atomic memories as well as based on diamond color centers \cite{Acosta2013}, trapped ions \cite{Staanum2002} or rare-earth ions doped solids \cite{Bartholomew2018,Chaneliere2015}, whenever ac-Stark modulation is feasible. While the presented SSM allows one-dimensional phase modulation, an extension to a subset of 2D phase profiles is straightforward by using two orthogonally placed 1D SSM modules, albeit such 2D profiles must be composed as an outer sum of 1D profiles.
 
In this work we demonstrate SSM modulation capabilities in the continuous domain as well as perform a proof-of-principle demonstration of a conditional SSM modulation, albeit without active feedback.
In particular, SSM is utilized to compensate a parabolic 1D phase introduced by a cylindrical lens placed in the readout path of the memory, enabling simple, indirect far-field characterization of the SSM modulation. 
Additionally, we perform a direct, spatially resolved, interferometric near-field characterization of the spin-wave phase profile introduced by SSM. In the measurement, interferograms are collected on an intensified sCMOS camera \cite{Lipka2018} after interfering the readout pulse with a weak signal beam which is also utilized for memory write-in. Near-field imaging of the memory readout let us also study the spin-wave decoherence due to the spatial intensity noise of the SSM beam, which is the main factor limiting SSM performance for large phase shifts.

In the conditional modulation protocol we store a light pulse as a coherent spin-wave state and split it in the temporal domain. The first part is readout as a coherent light state which could be measured and provide feed-back information to engineer SSM modulation of the remaining spin-wave state, shaping the subsequent readouts. 
We utilize the protocol without feedback to demonstrate phase coherence between two temporally separated memory readouts. The second readout is angularly shifted via a saw-tooth SSM modulation and the near-field interferograms of both readouts are collected on a single I-sCMOS camera frame. 

\begin{figure*}[htbp]
\centering
\fbox{\includegraphics[width=\linewidth]{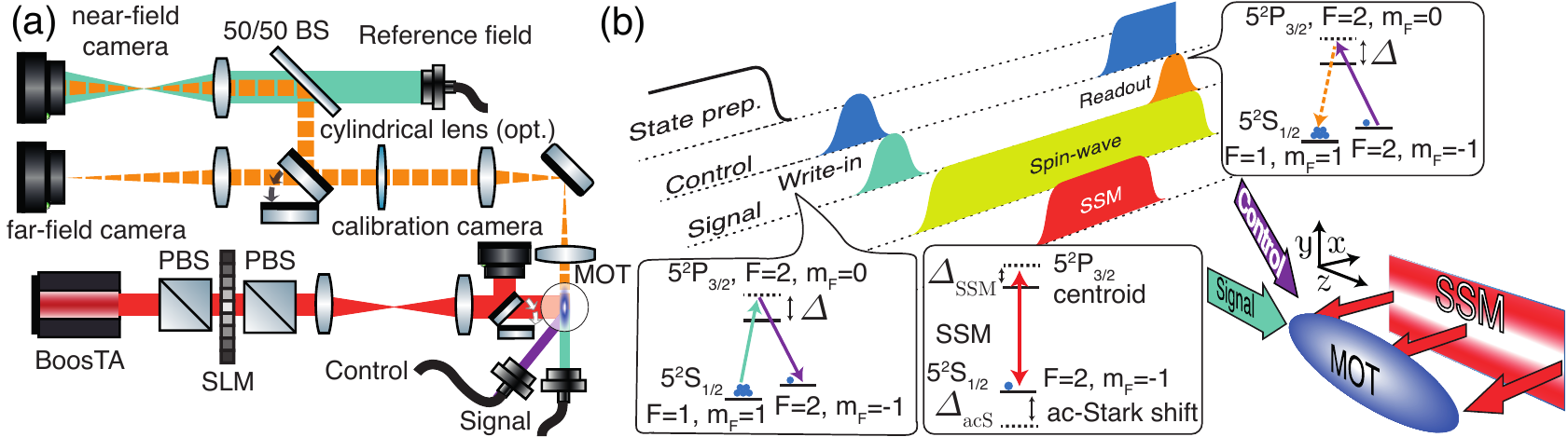}}
\caption{(a) Simplified experimental setup. Spatial spin-wave modulator (SSM) consists of a tapered laser amplifier (BoosTA) and a spatial light modulator (SLM) configured for SSM beam intensity profile shaping and imaged onto the atomic ensemble {\textendash} the heart of a multimode quantum memory. (b) Memory and SSM sequence. Spatial intensity profile of the SSM beam induces spatially varying ac-Stark shift producing phase modulation of a stored spin-wave state, retained in the readout light pulse. }
\label{fig:setup}
\end{figure*}
\section{Light-atom interface}
To demonstrate SSM modulation we employ a wavevector-multiplexed quantum memory \cite{Parniak2017} realized in a cold $^{87}\mathrm{Rb}$ ensemble trapped in a magneto-optical trap (MOT). Fig.~\ref{fig:setup} conceptually depicts the experimental setup and operation sequence for the memory and SSM. The memory itself operates in a lambda system depicted in Fig.~\ref{fig:setup} (b), with a strong control beam $\mathcal{E}_c$ detuned $\Delta = 20\;\mathrm{MHz}$ from the $|h\rangle=5^2\mathrm{S}_{1/2},\;F=2,\;m_F=1 \rightarrow |e\rangle= 5^2\mathrm{P}_{3/2},\;F'=2,\;m_F=0$ transition and a weak signal beam $\mathcal{E}_s$ two-photon resonant with $|h\rangle \rightarrow |g\rangle=5^2\mathrm{S}_{1/2},\;F=2,\;m_F=-1$. The atoms are initially prepared in the $|g\rangle$ state. The control beam is split and a part is electro-optically modulated at $6.8\;\mathrm{GHz}$. A modulation side-band is isolated via an actively stabilized Fabry-P\'erot etalon to obtain the signal beam \cite{Parniak2017}. Cascade offset locking ensures frequency stability between control, optical pumping and cooling lasers \cite{Lipka2017,Parniak2017}.

Via ca. $300\;\mathrm{ns}$ coincident control and signal beam pulses the signal beam is mapped onto a coherence $\rho_{gh}(x,y,z)\propto \mathcal{E}_s^* \mathcal{E}_c \exp(i\Delta_0 t)$ between the $|g\rangle$ and $|h\rangle$ ground states, where $\Delta_0=2\pi\times6.8\;\mathrm{GHz}$ corresponds to their hyperfine structure energy level separation. In the process, a coherent spin-wave state is created ${\sf S}(\mathbf{r}) = \mathcal{N}(\mathbf{r}) \rho_{gh}(\mathbf{r})\exp(-i\Delta_0 t)$, where $\mathcal{N}(\mathbf{r})$ corresponds to the atomic density and $\mathbf{r}=(x,y,z)$. Spatial dependence of the spin-wave is encompassed in its wavevector $\mathbf{K_\mathrm{sw}}=\mathbf{K_c}-\mathbf{K_s}$ with $\mathbf{K_c}$ and $\mathbf{K_s}$ corresponding to the wavevectors of the control and signal beam respectively. Subsequent application of a control beam pulse converts the stored spin-wave ${\sf S}$ state into a readout pulse conserving its transverse phase profile. 

\section{Ac-Stark spatial spin-wave modulator}
The spatial spin-wave modulator (SSM) relies on a spatially varying ac-Stark effect obtained with a strong SSM beam, detuned $\Delta_\mathrm{SSM}\approx1.5\;\mathrm{GHz}$ as calculated from the $|h\rangle$ to $5^2\mathrm{P}_{3/2}$ centroid. SSM beam intensity profile $I(y)$ is shaped by a spatial light modulator (Holoeye Pluto HES 6010) configured for amplitude modulation, as depicted in Fig~\ref{fig:setup} (a). The beam intensity $I(y)$ induces spatially varying ac-Stark shift $\Delta_\mathrm{acS}$ of the energy difference between $|g\rangle$ and $|h\rangle$ inherently changing the phase accumulation rate of the stored coherence $\rho_{gh}$. This way, by a proper adjustment of the SLM pattern, beam intensity and pulse duration $T$, 1D transverse spatial phase $\varphi(y)\propto I(y)T$ \cite{DeEchaniz2008} of a stored spin-wave state ${\sf S}$ can be engineered ${\sf S}(\mathbf{r})\rightarrow {\sf S}(\mathbf{r})\exp[i \varphi(y)]$. 

Note that modulation in the $z$ direction is also possible in this configuration. 

\begin{figure}[htbp]
\centering
\fbox{\includegraphics[width=\linewidth]{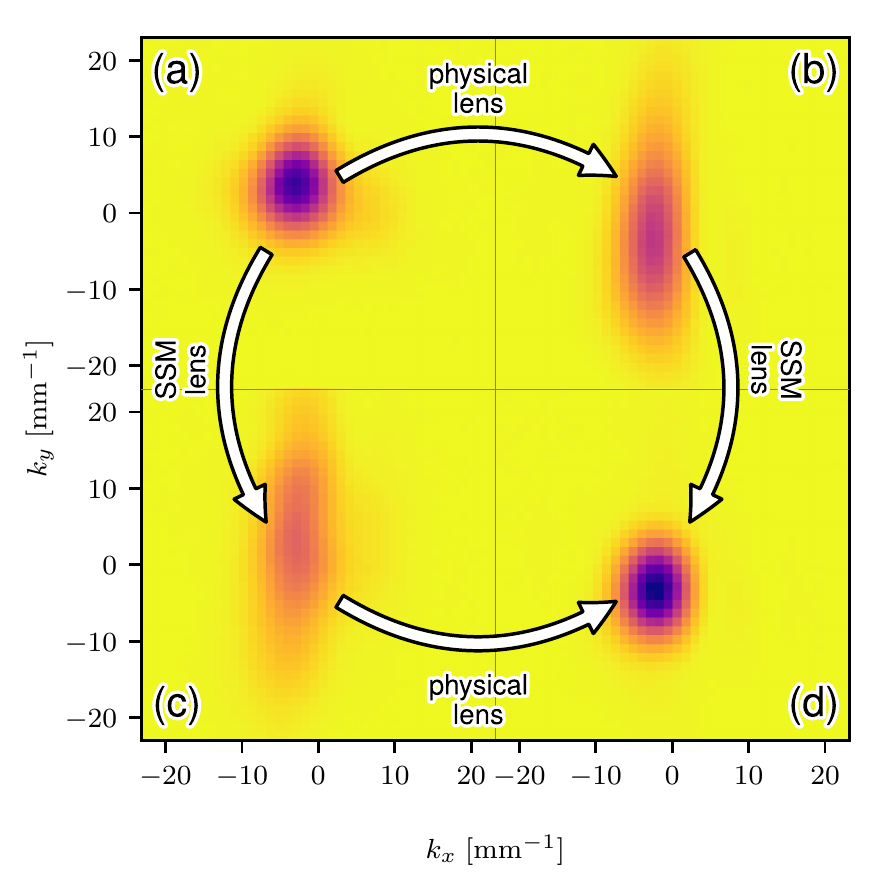}}
\caption{Far-field image of the memory readout (a) unaltered, (b) with a cylindrical lens introduced in the readout path. (c) SSM imposes 1D parabolic phase profile onto a stored spin-wave (d) compensating the cylindrical lens upon memory readout with efficiency of $\eta\approx80\;\%$ in terms of preserved total energy of the unaltered readout. The spatial fidelity between unaltered $I_0(x',y')$ and compensated $I(x',y')$ readout intensities amounts to $\mathscr{F}\approx96\%$.}  
\label{fig:circle}
\end{figure}
\begin{figure}[htbp]
\centering
\fbox{\includegraphics[width=\linewidth]{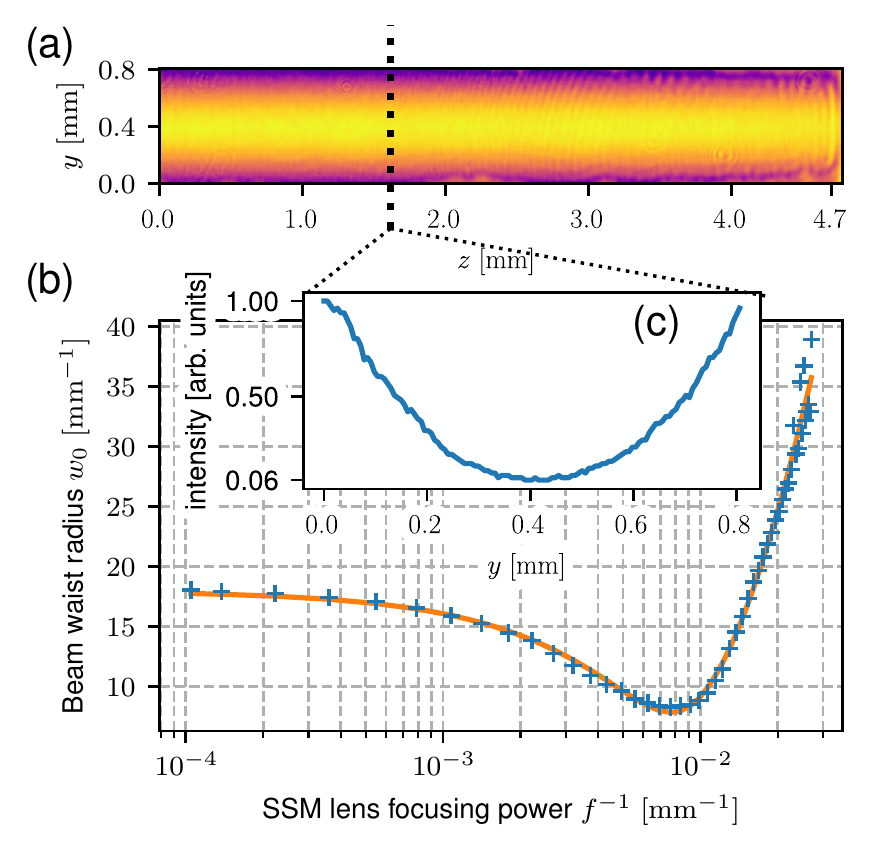}}
\caption{SSM compensation of a cylindrical lens introduced in the readout path of the quantum memory. (a),(c) SSM beam intensity profile parabolic in $y$ and flat in $z$ direction, as registered by a calibration camera placed at a position optically equivalent to the MOT location.  (b) SSM lens focusing power proportional to the SSM beam intensity and determined from the readout beam waist radius $w_0$ in $y$ direction, in the far-field. Solid line corresponds to a fitted theoretical model (vide main text). }
\label{fig:ff}
\end{figure}
\section{Far-field characterization}
We shall now harness the flexibility of SSM modulation to demonstrate shaping of the intensity profile of the memory readout in the far-field. For a proof-of-principle demonstration, we shall introduce a cylindrical lens in the readout route of the memory and refocus the readout by imposing a phase profile on a spin-wave state while it is still stored in the memory. With a simple feedback the required phase modulation could be adjusted to compensate intrinsic optical setup aberrations. Furthermore, analogous SSM lens phase profiles in 2D could be employed to dynamically switch between the near-field and far-field measurements of the memory readout.

A basic idea of SSM compensation is illustrated in Fig.~\ref{fig:circle}. The memory readout is observed in the far-field ($f_\mathrm{eff} = 50\;\mathrm{mm}$) of the atomic cloud on a camera (Basler sca1400-17fm). A cylindrical lens ($f_\mathrm{ph} = -2000\;\mathrm{mm}$ ) compound of tilted plano-convex ($f = +500\;\mathrm{mm}$) and meniscus ($f = -400\;\mathrm{mm}$) lenses is introduced in the near-field of the atomic ensemble. Due to magnification $M=4$ of the imaging setup, the effective focal length at the ensemble would appear $M^2=16$ times smaller.  
With SSM we impose a cylindrical phase profile $\varphi_{\pm}(y)=ky^2/(2f_\pm)$ onto a coherent spin-wave state stored in the memory. Here $k=2\pi/\lambda$, $\lambda=780\;\mathrm{nm}$. 

The sign of an SSM lens can be controlled twofold. On one hand, a positive (maximal intensity at the center) or a negative (maximal at the borders) curvature of the SSM beam intensity profile can be employed directly. On the other hand, for SSM beam detuning $\Delta_\mathrm{SSM}$ much higher than the natural line width, the ac-Stark shift $\Delta_{\mathrm{acS}}$ [vide energy level diagram in Fig. \ref{fig:setup}(b)] is inversely proportional to $\Delta_\mathrm{SSM}$ and their signs are opposite \cite{DeEchaniz2008,Parniak2018}. Additionally, the phase-accumulation rate of the stored spin-wave is generally proportional to $\Delta_\text{acS}$. Therefore, switching $\Delta_\mathrm{SSM}$ from blue-detuned to red-detuned alters the ac-Stark shift $\Delta_\text{acS}$ from red-shifted to blue-shifted and consequently reverses the sign of any further imposed phase.
 
Altering the total energy of the SSM pulse by controlling its power or duration yields precise control on the magnitude of the SSM phase modulation. Conversely, with the SSM cylindrical lens phase profile, its focal length can be continuously varied. \newline
\indent With the additional physical lens introduced in the setup and assuming Gaussian spatial distribution of the spin-wave amplitude, we may write the far-field amplitude of the readout beam as
\begin{equation}
\begin{aligned}
\mathcal{A}(x',y')\propto  \mathcal{F}&  \Bigg[  \exp\bigg\{-\frac{y^2}{w^2_\mathrm{sw}}-\Gamma(\varphi)\\&+i\left[\varphi_{\pm}(y)+M^2\varphi_\mathrm{ph}(y)\right]\bigg\}\Bigg](y'\frac{|\mathbf{K}|}{f_\mathrm{eff}}),
\end{aligned}
\label{eq:aff}
\end{equation}
where $w_\mathrm{sw}$ is the spin-wave waist radius, $\varphi_\mathrm{ph}(y)$ is the phase introduced by the physical lens and $f_\mathrm{eff}$ is an effective focal length of the imaging system. Additionally, we introduce $\Gamma(\phi)$ which corresponds to the total decoherence of the spin-wave, caused by the spatial inhomogeneities of the SSM beam intensity (for an SSM lens deviations from a parabolic intensity profile). At this point, let us make an assumption $\Gamma(\varphi)=\gamma\varphi^2$ which is described in detail in Sec. \ref{sec:Decoherence - SSM limitations}. The assumption is also consistent with an observation of how the total readout intensity $I=\int |\mathcal{A}(x',y')|^2\textrm{d}x'\textrm{d}y'$ scales with the SSM lens power. 

To characterize the quality of the SSM modulation, we shall consider the efficiency $\eta$ and the spatial fidelity $\mathscr{F}$ as two figures of merit. The former corresponds to the percentage of the total readout energy lost due to SSM modulation, while the later measures how closely the obtained readout amplitude resembles the desired one, neglecting the total intensity loss. For the far-field characterization, we shall consider the modulus of the readout amplitude either with SSM and the physical lens inserted $|\mathcal{A}(x',y')|=\sqrt{I(x',y')}$ or unaltered, without neither of those $|\mathcal{A}_0(x',y')|=\sqrt{I_0(x',y')}$.   We achieve the efficiency of $\eta = I/I_0 \approx 80\%$, while the fidelity given by a normalized scalar product:
\begin{equation}
\mathscr{F}=\frac{\langle|\mathcal{A}(x',y') \mathcal{A}_0(x',y')|\rangle_{x',y'}}{\sqrt{\langle I(x',y')\rangle_{x',y'}\langle I_0(x',y')\rangle_{x',y'}}}
\label{eq:fidelity}
\end{equation}
reaches $\mathscr{F}=96\%$. Observed intensities $I(x',y')$, $I_0(x',y')$ are depicted in Fig. \ref{fig:circle} (d) and Fig. \ref{fig:circle} (a) respectively.

Following Eq.~\ref{eq:aff}, the SSM lens focal length can be determined from the readout beam waist radius in the far-field $w_0$, as depicted in Fig.~\ref{fig:ff}, by fitting the model with $w_\mathrm{sw}$, $\gamma$, $f_\mathrm{ph}$ and a factor scaling imposed phase $\varphi_\pm(y)$ to the beam intensity $I(y)$ as free parameters. 
Employing $3\;\mu\mathrm{s}$ SSM modulation and a negative SSM beam intensity curvature, focal lengths down to ca. $f=\pm40\;\mathrm{mm}$ could be obtained. With two orthogonally placed 1D SSM modulators, such focal lengths would be sufficient for a dynamic far-field {\textendash} near-field switching of the memory readout. The active selection of the measurement basis is essential in free-space high-dimensional quantum communication based on Einstein-Podolsky-Rosen steering \cite{Dabrowski2017,Edgar2012,Aspden2013}. In such a protocol the conjugate position {\textendash} momentum bases correspond to the near-field and far-field respectively.

\section{Interferometric characterization}
To precisely characterize the SSM modulation, we resort to a spatially resolved interferometric measurement of the memory readout. The readout pulse interferes with a weak signal beam {\textendash} a reference field previously employed in the memory write-in yet now fiber routed as depicted in Fig.~\ref{fig:setup} (a), magnified and additionally tilted ca. $22\;\mathrm{mrad}$. Without SSM modulation, diagonal interference fringes are registered by a single-photon sensitive image intensified sCMOS camera placed in the near-field of the atomic ensemble (effective pixel pitch $3.25\;\mu\mathrm{m}$) and operated in the linear amplification regime. 

Registered interference pattern can be represented as $p(\mathbf{r_\bot})=h(\mathbf{r_\bot})\cos[\mathbf{K_0}\cdot\mathbf{r_\bot}+\Delta\varphi(\mathbf{r_\bot})+\varphi(\mathbf{r_\bot})]$, where $\mathbf{r_\bot}=(x,y)$ and $h(\mathbf{r_\bot})$ corresponds to the spatial intensity distribution of the readout, which is affected by the atomic density and spin-wave to light conversion efficiency, $\Delta\varphi(\mathbf{r_\bot})$ denotes residual spatial phase (e.g. caused by optical setup misalignment or aberrations) while $\varphi(\mathbf{r_\bot})$ represents the phase imposed by SSM. Due to the constant linear phase $\mathbf{K_0}\cdot\mathbf{r_\bot}$, any registered phase profile $\Delta\varphi(\mathbf{r_\bot})+\varphi(\mathbf{r_\bot})$ remains well separated from the zero-frequency component in the Fourier domain and can be easily isolated via a rectangular filter \cite{Bone86}. Denoting $q(\mathbf{r_\bot})=\exp[i\Delta\varphi(\mathbf{r_\bot})+i\varphi(\mathbf{r_\bot})]$ and its Fourier transform as $Q(\mathbf{K})$, the Fourier transform of $p$ consists of two components $P(\mathbf{K})=H(\mathbf{K})*[Q(\mathbf{K-K_0})+Q^*(\mathbf{-K-K_0})]$, where $*$ denotes convolution and $H(\mathbf{K)}$ is the Fourier transform of $h(\mathbf{r_\bot})$. We select the filter window to contain only the first, positive frequency component. Subsequently applying the inverse Fourier transform results in an analytical signal $h(\mathbf{r_\bot}) \exp[i\mathbf{K_0\cdot r_\bot}+i\Delta\varphi(\mathbf{r_\bot})+i\varphi(\mathbf{r_\bot})]$ carrying the spatial phase information. At this stage, by performing subsequent measurements with and without SSM modulation, the linear carrier $\mathbf{K_0}\cdot\mathbf{r_\bot}$ and residual phase $\Delta\varphi(\mathbf{r_\bot})$ can be subtracted. The argument of the remaining expression yields $\varphi(\mathbf{r_\bot})$.

Note that in the far-field the readout pulse is focused to ca. $w_0=8.3\;\mathrm{mrad}$ beam waist radius; however, in the near-field we observe collective emission from the whole atomic ensemble making the near-field image relatively weak. Therefore, a camera with an image intensifier is required to obtain sufficient sensitivity while keeping the image large enough for a reasonable detection resolution.
With ca. $200$ frames per second, a single measurement lasts around $50\;\mathrm{s}$. As the employed interferometer involves a complex quantum memory setup constructed across two optical tables, on such time scales the interferometric phase drift becomes substantial. Therefore, simple averaging of collected camera frames is unfeasible. Instead, we perform a post-processing phase tracking and remove the global phase time dependence $\Phi(t)$ from the Fourier transformed and filtered frames. These can be further averaged before applying the inverse Fourier transform. To perform the phase tracking, we calculate the complex dot product of the subsequent Fourier filtered frames with the first ($t_0$) gathered {\textendash} reference frame $\langle Q(\mathbf{K},t)|Q(\mathbf{K},t_0)\rangle$, where now $\varphi(\mathbf{r_\bot},t)=\varphi(\mathbf{r_\bot})+\Phi(t)$. The argument of such a product corresponds to the relative global phase $\Phi(t)$ of subsequent frames with $\Phi(t_0)=0$.

\begin{figure*}[htbp]
\centering
\fbox{\includegraphics[width=\linewidth]{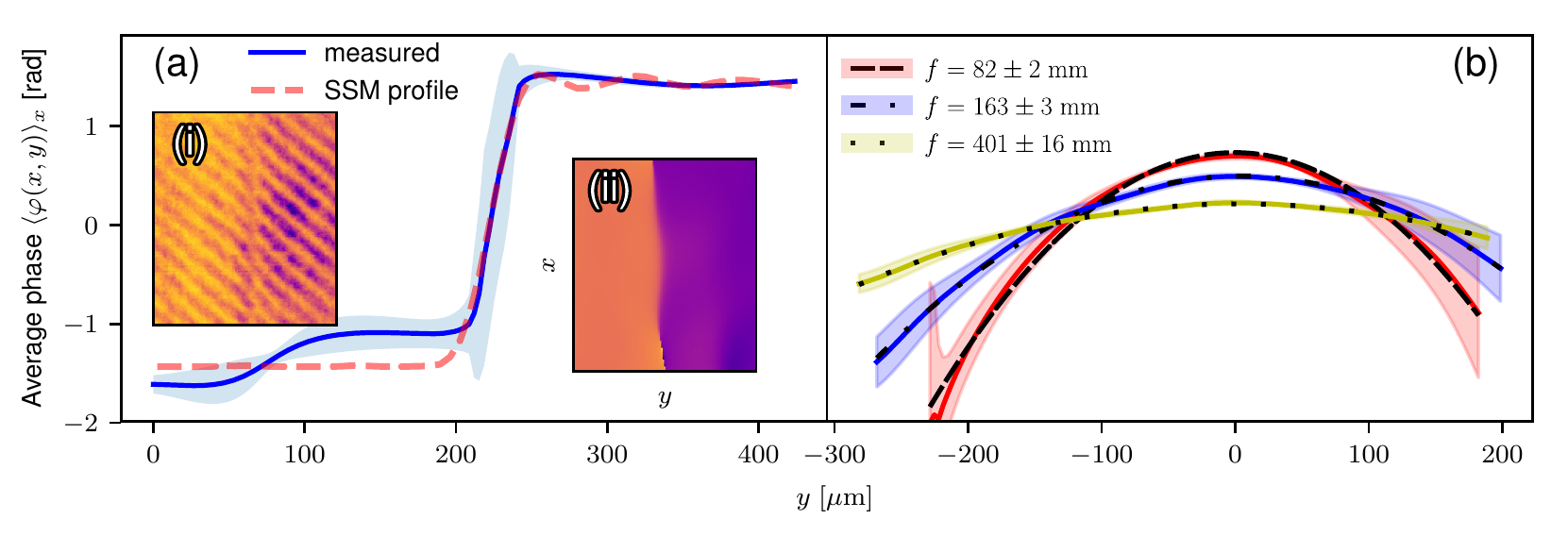}}
\caption{Interferometric measurement of a spatial 1D phase introduced by SSM modulation onto a coherent spin-wave state. (i) Exemplary observed interference pattern with (ii) corresponding reconstructed spatial phase.  (a) Step phase profile averaged over $x$ as measured (blue line) and expected from the intensity profile of the modulating SSM beam  (red dashed line). (b) SSM parabolic phase profile enables cylindrical lenses to be applied directly to a stored spin-wave. Focal length can be continuously varied with SSM beam power or modulation time.  }
\label{fig:nf}
\end{figure*}

For a demonstration of the method capabilities and SSM performance, let us perform a step phase modulation with a flat profile along $x$ and a rapid phase change at a chosen $y$. As demonstrated in Fig.~\ref{fig:nf} (a), spatial phase of ca. a half of the spatial extent of the spin-wave state has been approximately $\pi${\textendash}shifted. An exemplary collected fringe pattern has been depicted in the inset (i) of Fig.~\ref{fig:nf} (a). To maintain reasonable signal to noise ratio, all presented near-field data correspond to a selected region-of-interest (ROI). We select the ROI to be a minimal rectangular region bounding 2 standard deviations of a Gaussian fitted to the image of the memory readout without SSM modulation or reference field.
Inset (ii) depicts the phase $\varphi(x,y)$ retrieved from $10^4$ experiment realizations. The solid blue line in the main plot represents the phase averaged over $x$, $\langle\varphi(x,y)\rangle_x$ while the blue shaded area corresponds to its standard deviation along $x$. The red dashed line represents the $x$ average of the actual SSM beam intensity profile as observed by an additional calibration camera, arbitrarily rescaled to match observed phase profiles. The spatial fidelity between the measured $\varphi(x,y)$ and the expected step phase profiles $\varphi_\mathrm{0}(x,y)$ is given by an expression analogous to Eq.~\ref{eq:fidelity} with $\mathcal{A}\rightarrow\varphi$ and yields $\mathscr{F} = 98\%$. Efficiency in terms of the total preserved energy of the SSM modulated versus unaltered readout was $\eta=77\%$.

Panel (b) of Fig.~\ref{fig:nf} depicts the phase profiles of several cylindrical SSM lenses, averaged over the unaltered $x$ dimension. The shaded area corresponds to one standard deviation along $x$. A parabolic plane has been fitted to the retrieved phase $\varphi(x,y)$ to obtain the focal lengths and the $x$ average of the estimator has been depicted as the dashed, dash-dotted and dotted lines. Uncertainty in the focal length estimation has been obtained as a standard deviation across 10 independent measurements. The spatial fidelity ranges from $\mathscr{F}=95\%$ for $f=82\;\text{mm}$ to $\mathscr{F}=98\%$ for $f=163\;\text{mm}$ and $f=401\;\text{mm}$. For all focal lengths efficiency remained above $\eta=80\%$.
\section{Decoherence - SSM limitations}
\label{sec:Decoherence - SSM limitations}
\begin{figure}[htbp]
\centering
\fbox{\includegraphics[width=\linewidth]{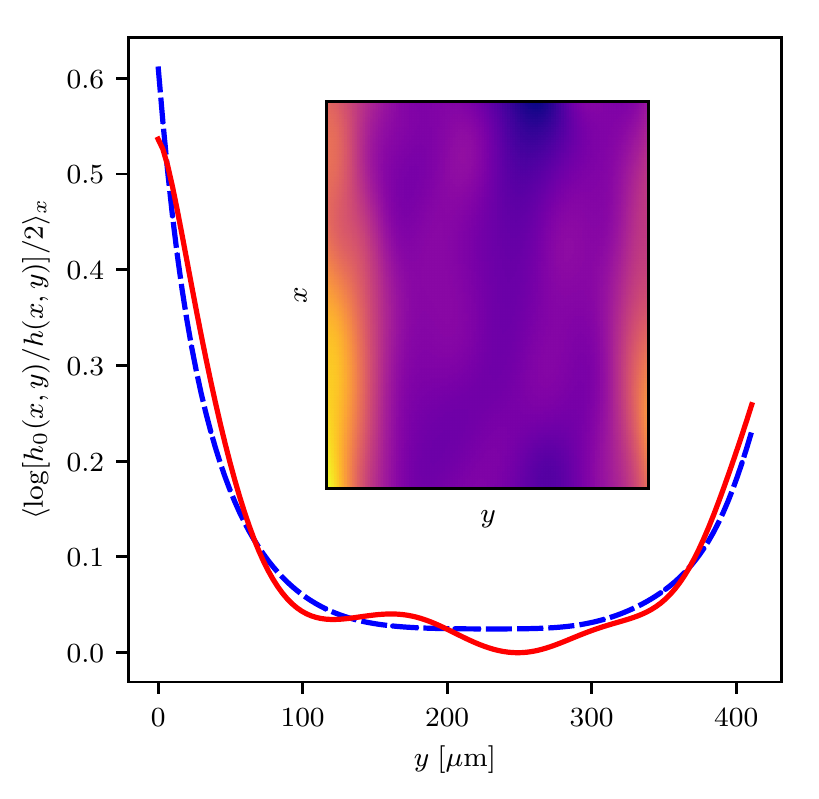}}
\caption{Spin-wave decoherence due to the spatial intensity noise of the SSM beam, reconstructed from the memory readout. The inset portrays a map of the SSM efficiency $h(x,y)/h_0(x,y)$
determined from interferometric near-field patterns of the memory readout with ($h(x,y)$) and without ($h_0(x,y)$) SSM modulation. Main plot depicts the $x$ average of the decoherence factor (red line) and its proportionality ($\gamma$) to the square of the SSM imposed phase $\varphi(y)^2$  (dashed line).} 
\label{fig:decoh}
\end{figure}
Due to experimental imperfections, SSM modulation degrades the efficiency of the spin-wave to light conversion during memory readout. With a far-detuned SSM beam the main factor contributing to the spin-wave decoherence are the inhomogeneities of the SSM beam intensity which may be introduced by scattering on the SLM, parasitic interference fringes and optical setup imperfections. Intensity inhomogeneities transfer to imposed phase degrading the phase matching of the readout.

Let us begin by observing the SSM beam on the calibration camera, situated in a position optically equivalent to that of the atomic cloud. We shall fix $y$ and focus on a single line of pixels along $z$. Ideally, all pixels should have the same value, corresponding to some beam intensity $I_0(y)$; however, imperfections introduce a $z$ dependent component so the total intensity becomes $I(y,z)$ = $I_0(y)+\Delta I(z)$. To remain independent of the selected SSM pattern, let us treat a line of pixels as a statistical ensemble. We have observed that a typical distribution of pixel values is Gaussian with a standard deviation relative to the mean of ca. $6\%$ for high intensity pixel lines independent of the intensity and growing for very dim lines up to $30\%$. Furthermore, the length of the correlation between pixel values is much smaller (ca. $37\;\mu\mathrm{m}$ at $-3\;\mathrm{dB}$) than the effective MOT length $L\approx1\;\mathrm{cm}$. Therefore, let us treat $\Delta I(z)$ for each $z$ as identically distributed independent zero-mean Gaussian random variables (or equivalently white Gaussian noise) with probability of obtaining a given $\Delta I$ value given by $P(\Delta I(z) = \Delta I)=1/(\sigma \sqrt{2\pi})\exp[-\Delta I^2/(2\sigma^2)]$.
With this assumption we can look at the memory readout amplitude $\mathcal{A}(x,y)$ after imposing SSM spatial phase $\varphi(y,z)=\alpha I(y,z) T$ onto the stored spin-wave, where $T$ corresponds to the interaction time and $\alpha$ denotes the proportionality factor. Neglecting the $z$ dependence of the atomic cloud density, the readout amplitude is given by:
\begin{equation}
\begin{aligned}
|\mathcal{A}(x,y)|&\propto\left|\frac{1}{L}\int_0^L \mathrm{d}z \exp\lbrace i\alpha T [I_0(y)+\Delta I(z)]\rbrace\right|=\\
 &=\left|\int_{-\infty}^{\infty} \mathrm{d}(\Delta I)P(\Delta I) \exp(i\alpha T \Delta I)\right|=\\
 &=\exp(-\frac{1}{2}\alpha^2 T^2 \sigma^2),
\end{aligned}
\end{equation}

Note that while the imposed SSM phase is proportional to the beam intensity $\varphi(y,z)=\alpha T I(y,z)$, the noise variance scales quadratically $\sigma^2 \propto I_0^2$.
Therefore, the SSM efficiency can be recasted to $\exp[-\Gamma(\varphi)]$ as in Eq.~\ref{eq:aff}, with $\Gamma(\varphi)=\gamma\varphi(y)^2$. 
This quadratic $\Gamma(\varphi)$ dependence is directly confirmed by a near-field interferometric characterization of the SSM modulated readout. Such measurement enables direct observation of the decoherence with a spatial resolution. By applying the Fourier domain filtering we obtained an analytical signal carrying the phase information in its argument as well as the spatial readout intensity in its modulus $h(x,y)$. Comparing the memory readouts with ($h(x,y)$) and without ($h_0(x,y)$) SSM modulation, the decoherence factor can be obtained $\log[h_0(x,y)/h(x,y)]/2$. Employing $h(x,y)$ instead of raw readout images mitigates the influence of the image noise and removes the constant background.
Fig.~\ref{fig:decoh} depicts the proportionality $\gamma=(4.2\pm1.4)\times10^{-2}$ of the $x$ average of the decoherence factor to the square of the imposed phase $\varphi(y)^2$ obtained as a fit to the raw phase retrieved from an interferometric measurement. The employed SSM modulation corresponds to an SSM lens ($f=82\pm2\;\mathrm{mm}$) depicted in Fig.~\ref{fig:nf} (b). Measured $\gamma$ factor gives a relative standard deviation of the intensity $(\langle\Delta I^2\rangle)^{1/2}/I_0 = (2\gamma)^{1/2} = (29\pm5)\%$. It is substantially higher than the pixel values inhomogeneities (ca. $6\%$) observed with the calibration camera. Virtually, with the camera we should be able to observe all deviations from the desired intensity pattern; however, in practice limited camera resolution and precision of its positioning hinders our ability to observe certain kinds of distortions, such as very dense fringes which may contribute to the interferometrically measured $29\%$.

Ideally, the intensity distribution of the SSM beam in terms of the probability density function would approach a delta function centered at the average intensity. As the incoherent light could be used for SSM modulation recent methods of engineering speckle intensity distributions \cite{Bender2018} may be perhaps utilized to improve SSM efficiency.

\section{Phase stability of the memory}
A coherent spin-wave state can be readout as several temporally separated light pulses by employing a train of short control field pulses. With the SSM modulation, it is possible to alter the spatial phase of the subsequent readouts. In principle, the modulation of $n$-th pulse could depend on the measurement outcomes on $1,\ldots,n-1$ previous pulses via a feedback loop. In this way, SSM could be employed to realize spatially resolved adaptive measurements, opening new possibilities in optical communication and information processing protocols. Inherently, quantum memory serving in a such a scheme would be required to preserve phase coherence between successive readouts. Here we operate the memory and employ SSM in an analogous scheme, albeit without feedback, to show phase coherence between readouts of a temporally split coherent spin-wave state.

\begin{figure*}[htbp]
\centering
\fbox{\includegraphics[width=\linewidth]{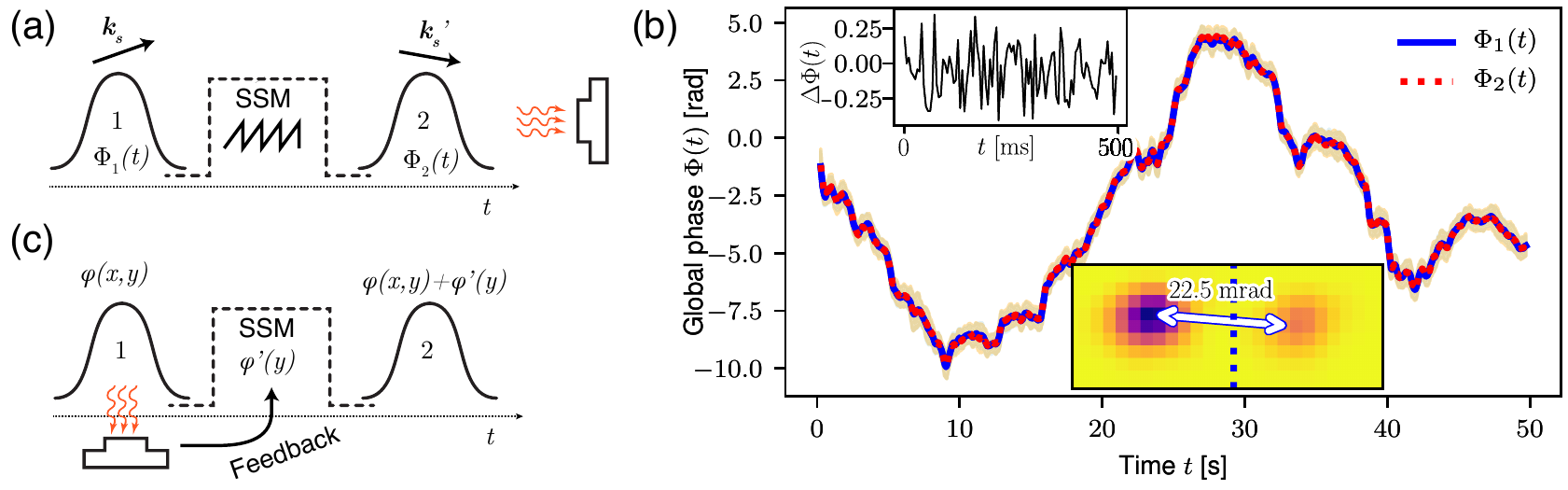}}
\caption{(a),(b) Measurement of the phase stability between two temporally separated partial readouts of a single coherent spin-wave state. (a) Memory operation and SSM sequence. Second readout is angularly shifted by imposing linear phase on the spin-wave with the SSM. Interference patterns of both readouts are collected on a single camera frame. (b) Global phase $\Phi_i(t)$ of the readouts $i=1,2$ tracked across $10^4$ frames. Upper inset depicts the phase difference $\Delta\Phi(t) = \Phi_1(t)-\Phi_2(t)$ for the first $500\;\mathrm{ms}$. Lower inset illustrates readouts separation in the Fourier domain. (c) SSM could be employed along with the temporal spin-wave splitting to realize adaptive measurements. The modulation profile $\varphi'(t)$ would be conditioned on the first readout measurement outcome.}
\label{fig:double}
\end{figure*}

The employed memory and SSM sequence has been depicted in Fig.~\ref{fig:double} (a). The first readout is performed after $1\;\mu\mathrm{s}$ memory time with a short $200\;\mathrm{ns}$ control field pulse. Further, SSM saw-tooth profile modulation shifts the transverse wave-vector of the remaining spin-wave state making the readouts spatially separated in the far-field. The modulation intensity and time is adjusted to meet the conditions of a Fresnel surface and obtain linear phase profile. Second readout takes place after $10\;\mu{s}$ memory time. Both readout pulses are interfered with the reference field and the fringe pattern is registered in the near-field on a single I-sCMOS camera frame.

In the Fourier domain parts of the registered interference pattern corresponding to the first and the second readout can be easily separated, as illustrated in the lower inset of Fig.~\ref{fig:double} (b). The vertical dashed line indicates the boundary selected to separate the first readout (left) from the SSM shifted second one (right). For each readout $i=1,2$ we independently track its global phase $\Phi_i(t)$ across $10^4$ gathered frames. As depicted in Fig.~\ref{fig:double} (b), the global phases $\Phi_1(t)$, $\Phi_2(t)$ well overlap on the scales of the interferometric phase drift. Here for readability rolling average with a window of $50$ frames ($250\;\mathrm{ms}$) has been applied. Filled area around the curves corresponds to one standard deviation in each rolling average window. The upper inset depicts the phase difference $\Delta\Phi(t) = \Phi_1(t)-\Phi_2(t)$ for the first gathered $500\;\mathrm{ms}$ without any averaging. Standard deviation of the phase difference $[\mathrm{Var}\Delta\Phi(t)]^{1/2} \approx 0.2\;\mathrm{rad}$ certifies phase stability between subsequent memory readouts.

The residual phase deviation $[\mathrm{Var}\Delta\Phi(t)]^{1/2}$ may originate from the phase deviation of the signal beam relative to the control beam, across the period between the two readouts of ca. $9\;\mu\mathrm{s}$. Such a deviation may be caused by a slow response of the feedback loop employed to actively stabilize the Fabry-P\'erot etalon which isolates the $6.8\;\mathrm{GHz}$ modulation sideband of the control beam to produce the signal beam. With the loop closed, we measured the power of the phase error signal to be flat $-63\;\mathrm{dBm}$ in the relevant frequencies range between $100\;\mathrm{kHz}$ and $10\;\mathrm{MHz}$. For the measurement a spectrum analyzer at the resolution band-width (RBW) of $1\;\mathrm{kHz}$ was employed. Due to $300\;\mathrm{ns}$ write pulse duration any noise is effectively low-pass filtered with a pole at ca $3.3\;\mathrm{MHz}$. Therefore, the measured phase error power density was filtered and integrated in the relevant frequencies range corresponding roughly to the time periods above the write pulse duration and below the readouts separation of $9\;\mu\mathrm{s}$. With the etalon free spectral range of $10\;\mathrm{GHz}$ and finesse of $100$, the phase of the transmitted field deviates at an rate of $1\;\mathrm{rad}/(50\;\mathrm{MHz})$ with the detuning from the transmission maximum, in the linear regime. With the phase error signal scaling of ca. $200\;\mathrm{mV}/\mathrm{rad}$ and accounting for additional $-30\;\mathrm{dB}$ from RBW, we obtain the RMS deviation of ca. $0.1\;\mathrm{rad}$ for $[\mathrm{Var}\Delta\Phi(t)]^{1/2}\approx[2\mathrm{Var}\Phi(t)]^{1/2}$. This result, together with possible phase noise contribution from $10\;\mathrm{m}$ fibers carrying the signal and reference beams, justifies the observed value of the residual phase deviation of ca. $0.2\;\mathrm{rad}$.

\section{Conclusions}
In this work we have brought the ac-Stark spin-wave modulation technique to the continuous domain, achieving modulation flexibility in 1D analogous to SLM modulators for light, yet with the advantage of working in the matter domain with long interaction time. Both far-field and a direct, interferometric near-field characterization of the spatial spin-wave modulator (SSM) reveals excellent spatial fidelity of the imposed phase profiles above $\mathscr{F}=95\%$. Spin-wave decoherence due to the SSM beam spatial intensity inhomogeneities, scaling with the square of the imposed phase is found to be the main factor limiting the SSM performance for large phase shifts. Nevertheless, for demonstrated manipulations relatively high efficiency around $\eta=80\%$ was maintained.

We have also proposed a scheme to perform conditional SSM modulation with a temporally split readout of a coherent spin-wave state. Such a scheme could be applied in adaptive measurements. Finally, we employ an analogous scheme without feedback to demonstrate that employed quantum memory retains good phase stability between subsequent partial readouts of a coherent spin-wave state.

With the demonstrated spatial fidelity and efficiency SSM should be readily feasible for high-fidelity single-photon spatial modulation. With the employed quantum memory, for single-photon states, additional spatial filtering is required \cite{Parniak2017}; however, simple few-mode ac-Stark operations in this regime has been successfully demonstrated \cite{Parniak2018}. Further improvements of the spatial fidelity and efficiency of SSM could be achieved by employing a higher resolution SLM and improving the calibration optical setup. This way, relatively high SSM beam spatial intensity inhomogeneities (standard deviation of $29\%$) could be substantially reduced leading to major improvements in modulation efficiency and extending the range of feasible phase-shifts.

A broad family of SSM compatible quantum memories along with additional ac-Stark modulation possibilities \cite{Parniak2018,Mazelanik2018} including discrete inter-mode operations on spin-waves in wavevector as well as temporal domains render the high-fidelity ($\mathscr{F}\geq95\%$), flexible SSM modulation a valuable tool for practical implementations of novel quantum information and communication protocols.

\begin{acknowledgments}
This work has been funded by National Science Centre, Poland (NCN) (Grants No. 2016/21/B/ST2/02559, 2017/25/N/ST2/01163, 2017/25/N/ST2/00713), Polish MNiSW "Diamentowy Grant" project no. DI2016 014846 and by the project ``Quantum Optical Communication System" carried out within the TEAM programme of the Foundation for Polish Science co-financed by the European Union under the European Regional Development Fund.
We would like to also thank Konrad Banaszek for the generous support.
\end{acknowledgments}

\bibliography{swlensbib}

\end{document}